\documentclass[aps,prd,reprint,twocolumn,superscriptaddress,showpacs]{revtex4-1}
\usepackage{graphicx}
\usepackage{mathrsfs}
\usepackage{bm}
\usepackage{amsmath}
\usepackage{dcolumn}
\usepackage{epstopdf}
\usepackage{dsfont}
\usepackage{amssymb}
\usepackage{tabularx}
\usepackage{array}
\usepackage{float}
\usepackage{color}
\usepackage{epstopdf}
\usepackage{mathrsfs}
\usepackage[colorlinks, linkcolor=blue,anchorcolor=blue,citecolor=blue,urlcolor=blue]{hyperref}

\begin{document}

%Title of paper
\title{Transverse shift in crossed Andreev reflection}

\author{Ying Liu}\email{ying\_liu@mymail.sutd.edu.sg}
\affiliation{Research Laboratory for Quantum Materials, Singapore University of Technology and Design, Singapore 487372, Singapore}

\author{Zhi-Ming Yu}\email{zhiming\_yu@sutd.edu.sg}
\affiliation{Research Laboratory for Quantum Materials, Singapore University of Technology and Design, Singapore 487372, Singapore}

\author{Jie Liu}
\affiliation{Applied Physics Department, School of Science, Xi'an Jiaotong University, Xi'an 710049, China}

\author{Hua Jiang}
\affiliation{School of Physical Science and Technology, Soochow University, Suzhou 215006, China}
\affiliation{Institute for Advanced Study, Soochow University, Suzhou 215006, China}

\author{Shengyuan A. Yang}
\affiliation{Research Laboratory for Quantum Materials, Singapore University of Technology and Design, Singapore 487372, Singapore}
\affiliation{Center for Quantum Transport and Thermal Energy Science, School of Physics and Technology, Nanjing Normal University, Nanjing 210023, China}

\begin{abstract}
Crossed Andreev reflection (CAR) is an intriguing effect that occurs in a normal-superconductor-normal junction. In CAR, an incoming electron from one terminal is coherently scattered as an outgoing hole into the other terminal. Here, we reveal that there exists a transverse spatial shift in CAR, i.e., the plane of CAR for the outgoing hole may have a sizable transverse shift from the plane of incidence for the incoming electron. We explicitly demonstrate the effect in a model system based on Weyl semimetals. We further show that the effect is quite general and exists when the terminals have sizable spin-orbit coupling. In addition, we find that the corresponding shift in the elastic cotunneling process shows different behaviors, and it vanishes when the two terminals are identical. Based on these findings, we suggest possible experimental setups for detecting the effect, which may also offer an alternative method for probing CAR.
\end{abstract}

\maketitle
\section{Introduction}
Andreev reflection is a unique scattering process that occurs at an interface between a normal-metal (or a doped semiconductor) and a superconductor~\cite{Andreev1964}. During Andreev reflection, an incoming electron from the normal-metal (N) side is reflected as a hole at the interface, and the missing charge of $(-2e)$ is transferred into the superconductor (S) as a Cooper pair.

Remarkably, the electron-hole conversion process can also happen nonlocally, giving rise to an intriguing phenomenon known as crossed Andreev reflection (CAR)~\cite{Byers1995,Deutscher2000}. It appears in hybrid normal-superconductor-normal (NSN) structures, as schematically illustrated in Fig.~\ref{Fig1}. When the thickness of the S layer is smaller than or comparable to the superconducting coherence length, an electron incident from the first N terminal can form a Cooper pair in S with another electron from the second N terminal, thereby coherently transmitting a hole into the second N terminal and making a nonlocal charge transport.

CAR has been receiving considerable research interest, partly because it provides a solution for generating entanglement between electrons in spatially separated regions which is needed for quantum computation and quantum information applications~\cite{Recher2001,Lesovik2001}. So far, the experimental detection of CAR mostly relies on the nonlocal transport measurement of quantities such as the nonlocal voltage or conductance~\cite{Beckmann2004,Russo2005,Cadden-Zimansky2006}. However, such measurement is often complicated by another competing nonlocal process---the elastic cotunneling (EC)~\cite{Falci2001}, during which the incident electron directly tunnels to the second terminal, leading to an opposite contribution (with respect to CAR) to the  nonlocal signal. It has been shown that to the lowest order in the interface transmission, the contribution from EC exactly cancels that from CAR in the nonlocal conductance~\cite{Falci2001}; whereas for more transparent junctions, the EC contribution tends to be dominant~\cite{Kalenkov2007}. Consequently, the recent research has mainly been focused on finding ways to enhance the CAR contribution~\cite{Chtchelkatchev2003,Dong2003,Yamashita2003,Yeyati2007,Cayssol2008,Veldhorst2010,Reinthaler2013,Soori2017} and also on developing new methods to detect the CAR process~\cite{Hofstetter2009,Herrmann2010,Wei2010}.

In this paper, we investigate a different aspect of CAR. For an incident electron beam that undergoes CAR, the corresponding trajectory defines two planes: the plane of incidence for the incident electron and the plane of CAR for the outgoing hole, as illustrated in Fig.~\ref{Fig1}. It seems natural that the two planes should coincide, which was always implicitly assumed. Here we show that this is \emph{not} always the case---the plane of CAR can actually have a sizable \emph{transverse} spatial shift from the plane of incidence.

This work is motivated by recent studies that discovered electronic analogs~\cite{Jiang2015,Yang2015b,Jiang2016,Wang2017} of Imbert-Fedorov shift in geometric optics~\cite{Fedorov1955,Imbert1972} and especially by our recent findings on the transverse shift in the local Andreev reflection~\cite{Liu2017c,Yu2017}. Here, using the general quantum mechanical scattering approach, we explicitly demonstrate the possible existence of a transverse spatial shift in CAR. For particular cases where a rotational symmetry is preserved, we find that the result can be exactly reproduced via a symmetry argument. We analyze three concrete model systems. The first has the two N terminals consisting of a (doped) Weyl semimetal (WSM)~\cite{Wan2011,Murakami2007}, where the low-energy carriers are described by Weyl fermions. We attribute the large anomalous transverse shift to the strong spin/pseudospin-orbit coupling (SOC) that is inherent for Weyl fermions. However, the presence of Weyl fermions (or any band crossing) is not a necessary condition for the shift. We explicitly demonstrate this point using the second system, where the two N terminal are doped semiconductors with SOC. Furthermore, as the third example, we show that sizable transverse shift can persist for a $p$S$n$ junction, where the two terminals are semiconductors doped into $p$- and $n$-type. Such setup has the advantage that EC can be completely suppressed for a range of excitation energies, making the nonlocal transport entirely due to CAR. The transverse shift in CAR can lead to measurable local signals on the second N terminal. For certain cases like the second and the third model systems, the shift gives rise to surface charge accumulations that can be measured electrically as transverse voltage signals (for which EC has no contribution). Thus, our work not only reveals a fundamental physical effect, it may also offer a promising alternative method for detecting CAR in experiment.

\section{Model and  Approach}

\begin{figure}[t]
	\includegraphics[width=8 cm]{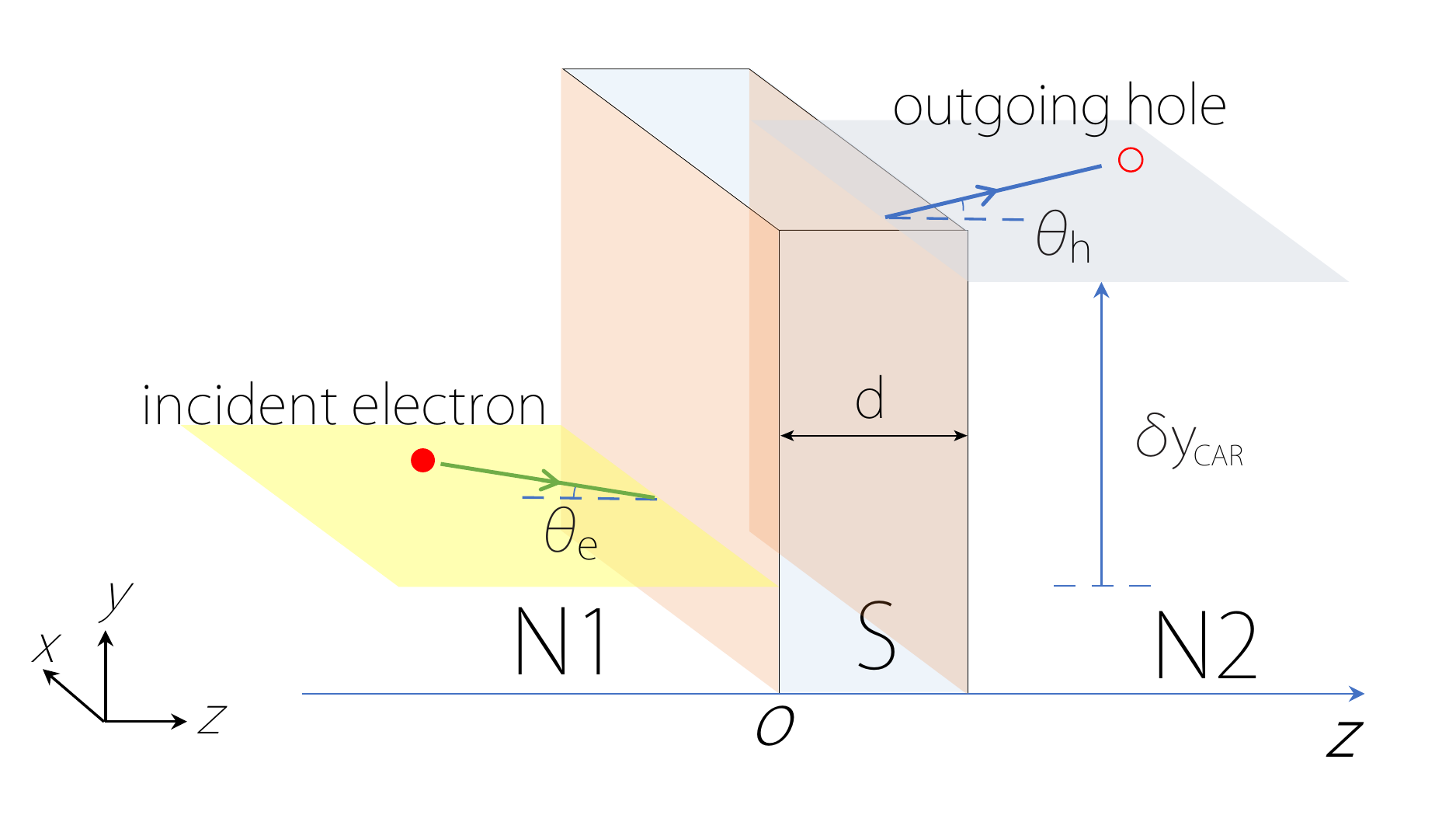}
	\caption{Schematic figure showing the transverse shift in CAR. In the hybrid NSN structure, an incident electron from terminal N1 is coherently scattered as an outgoing hole in terminal N2. There may exist a transverse shift $(\delta y_\mathrm{CAR})$ between the two scattering planes. }
	\label{Fig1}
\end{figure}

We consider the hybrid NSN structure as illustrated in Fig.~\ref{Fig1}. We assume that the system is extended in $x$ and $y$ directions (which amounts to saying that the system dimension in these two directions is much larger than the quasiparticle wavelength). The two NS interfaces are perpendicular to the $z$ direction, and are located at $z=0$ and $z=d$, respectively. The two terminals are denoted as N1 and N2. For studying the nonlocal scattering process, we take $d$ to be comparable to the superconducting coherence length $\xi$ for the S layer in the calculation.

The quasiparticle scattering properties in the structure are described by the microscopic Bogoliubov-de Gennes (BdG) equation~\cite{Gennes1966,Blonder1982}:
\begin{eqnarray}\label{BdG}
  \left[\begin{array}{cc}
    H_0+U(\bm r)-E_F&\Delta(\bm r)\\
    \Delta^*(\bm r)&-\mathcal{T}H_0\mathcal T^{-1}-U(\bm r)+E_F
  \end{array}\right]\psi=\varepsilon\psi.\nonumber\\
\end{eqnarray}
Here, $H_0$ is the electronic Hamiltonian in the normal state, $U(\bm r)$ represent a potential energy offset between the different regions, $E_F$ is the Fermi energy, $\mathcal T$ is the time-reversal operator, and the excitation energy $\varepsilon$ is measured from Fermi level. The wavefunction $\psi=(u,v)^T$ is a multicomponent spinor with $u$ $(v)$ standing for the electron (hole) state. In Sec.~III and IV, we shall assume that the two N terminals are of the identical material, hence they share the same $H_0$, and $U(\bm r)=-U_0[\Theta(z)-\Theta(z-d)]$ where $\Theta$ is the Heaviside step function.  The superconducting pair potential $\Delta(\bm r)$ is nonvanishing in the S region. Here, we take the usual step function model~\cite{Blonder1982} with $\Delta(\bm r)=\Delta_0[\Theta(z)-\Theta(z-d)]$, which has been shown to be a good approximation to the full self-consistent solution of the BdG equation for such hybrid structure~\cite{Hara1993,Plehn1994}. Particularly, it is accurate when there is large Fermi momentum mismatch across the interfaces (which effectively reduces the coupling between the layers)~\cite{Veldhorst2010}, which is the case that we are interested in. The mean-field requirement for superconductivity is that $E_F+U_0\gg \Delta_0$ in the S region, meaning that the Fermi wavelength in S should be much smaller than the coherence length. Meanwhile, the Fermi wavelength in N is not constrained to be small. Particularly, when N is of a doped semiconductor or semimetal, we may have $E_F$ comparable to $\Delta_0$, provided that $U_0$ is large.

The scattering properties for the quasiparticles are encoded in the scattering amplitudes, which are obtained by solving the scattering states for the BdG equation. The procedure is standard~\cite{Blonder1982}: one solves the eigenstates for each region (given by plane waves), and connects them at the interfaces using proper boundary conditions. For an incident electron plane wave state $\psi^{e+}$ from N1, there are four scattering processes: normal reflection (NR) as an electron into N1, Andreev reflection as a hole into N1, EC as an electron into N2, and CAR as a hole into N2. In this work, since the focus is on the transmitted quasiparticles in N2, we will mainly consider the transmission amplitudes $t_e$ and $t_h$ for EC and CAR processes.

Note that the spatial shift is well defined only for a laterally confined quasiparticle beam. In the standard treatment~\cite{Beenakker2009,Jiang2015,Liu2017c,Yu2017}, the beam $\bm\Psi$ is represented as a superposition of the partial waves. For example, the incident beam can be expressed as
\begin{eqnarray}\label{Psie}
\bm \Psi^{e+}(\bm r)=\int d\bm k'\ w(\bm k'-\bm k)\psi^{e+}_{\bm k'}(\bm r).
\end{eqnarray}
Here $w$ is the beam profile required to be peaked at an average wave-vector $\bm k$. The specific form of $w$ does not affect the final result of the spatial shift. In practice, one usually takes a Gaussian form for $w$, with $w(\bm q)=\Pi_i w_i(q_i)$, where $w_i(q_i)=(\sqrt{2\pi }W_i)^{-1}\exp[-q_i^2/(2W_i^2)]$, and $W_i$ is the width for the $i$th component. The scattering of the beam through the NSN structure can be studied by analyzing the scattering of each partial wave component $\psi^{e+}_{\bm k'}$, which are described by the scattering amplitudes. For example, the outgoing hole beam via CAR is given by
\begin{equation}\label{Psih}
\bm \Psi^{h+}(\bm r)=\int d\bm k'\ w(\bm k'-\bm k)t_{h}(\bm k')\psi^{h+}_{\bm k'}(\bm r),
\end{equation}
where $\psi^{h+}$ denotes the forward propagating hole eigenstate in N2. Then, the spatial shift is obtained by comparing the central positions of the two beams (which define the two scattering planes). The details of the approach can be found in the following section for the study of concrete models.

We have a few remarks before proceeding. First, if the plane of incidence is of the $x$-$z$ plane (as illustrated in Fig.~\ref{Fig1}), then the transverse shift that we are looking for will be along the $y$ direction. We note that for systems with non-negligible anisotropy in the $x$-$y$ plane, the transverse shift would actually depend on the orientation of the plane of incidence~\cite{Yu2017}.

Second, the quantum scattering approach that we adopt here is quite general. Unlike the semiclassical approach which requires that the potential variation is slow and smooth over the quasiparticle wavelength~\cite{Yang2015b}, the quantum scattering approach does not suffer from this constraint. Particularly, it applies for sharp interfaces and for the case when the N region is of a doped semiconductor or semimetal with a large Fermi wavelength~\cite{Beenakker2009,Jiang2015,Liu2017c}.

Third, the approach makes it apparent that the transverse shift results from a change of interference among the partial waves during scattering.
As observed from Eqs.~(\ref{Psie}) and (\ref{Psih}), a quasiparticle beam can be regarded as a superposition of the partial waves ($\psi_{\bm k'}^{e+}$ or $\psi_{\bm k'}^{h+}$), and its trajectory is determined by the interference between these partial waves. In the scattering, each partial wave $\psi_{\bm k'}^{e+}$ is scattered to $\psi_{\bm k'}^{h+}$ with the scattering amplitude $t_h$. When the scattering amplitudes are different for different partial waves, the interference pattern between the partial waves could be altered, leading to a change in the trajectory for the scattered beam. In the following section, we shall see that this happens when the quasiparticles in N have a strong SOC.

\section{Model I: WSM/S/WSM Junction}

In the first model, we consider that the two N terminals are identical and are made of a doped WSM. WSM is a type of topological material, in which the conduction and valence bands touch at isolated points in the momentum space called the Weyl points~\cite{Armitage2018}. The low-energy carriers around the Weyl points are described by the Weyl equation. Then for electrons near a Weyl point (at $\bm K_0$), the corresponding $H_0$ in Eq.~(\ref{BdG}) may be expressed as (set $\hbar=1$)
\begin{equation}\label{Weyl}
	H_0=-i\chi\sum_i v_i\sigma_i\partial_i,
\end{equation}
where $\chi=\pm$ stands for the chirality of the Weyl electrons, $\sigma$'s are Pauli matrices, $v$'s are the Fermi velocities, and the subscript $i$ denotes the three spatial dimensions. The $\sigma$ in the model may stand for the real spin or a kind of pseudospin. Here, for concreteness, we let it to be the real spin. The model intrinsically has a strong SOC, which is a crucial point that we will analyze later.
In addition, a Weyl semimetal requires the breaking of the inversion symmetry $\mathcal P$ or the time-reversal symmetry $\mathcal T$~\cite{Armitage2018}. In this model, we assume that $\mathcal P$ is broken while $\mathcal T$ is preserved, then each time-reversal pair of Weyl points will share the same chirality and be described by the same $H_0$.

For the ease of analytical calculation, we take for the S region the same $H_0$ in Eq.~(\ref{Weyl}). However, this region needs to be heavily doped (with large $U_0$) to satisfy the mean-field requirement for superconductivity, as we discussed in Sec.~II. Later, we shall see that this choice of heavily doped Weyl model for the S region actually does not affect the key result.

\begin{figure}[t]
	\includegraphics[width=8.4 cm]{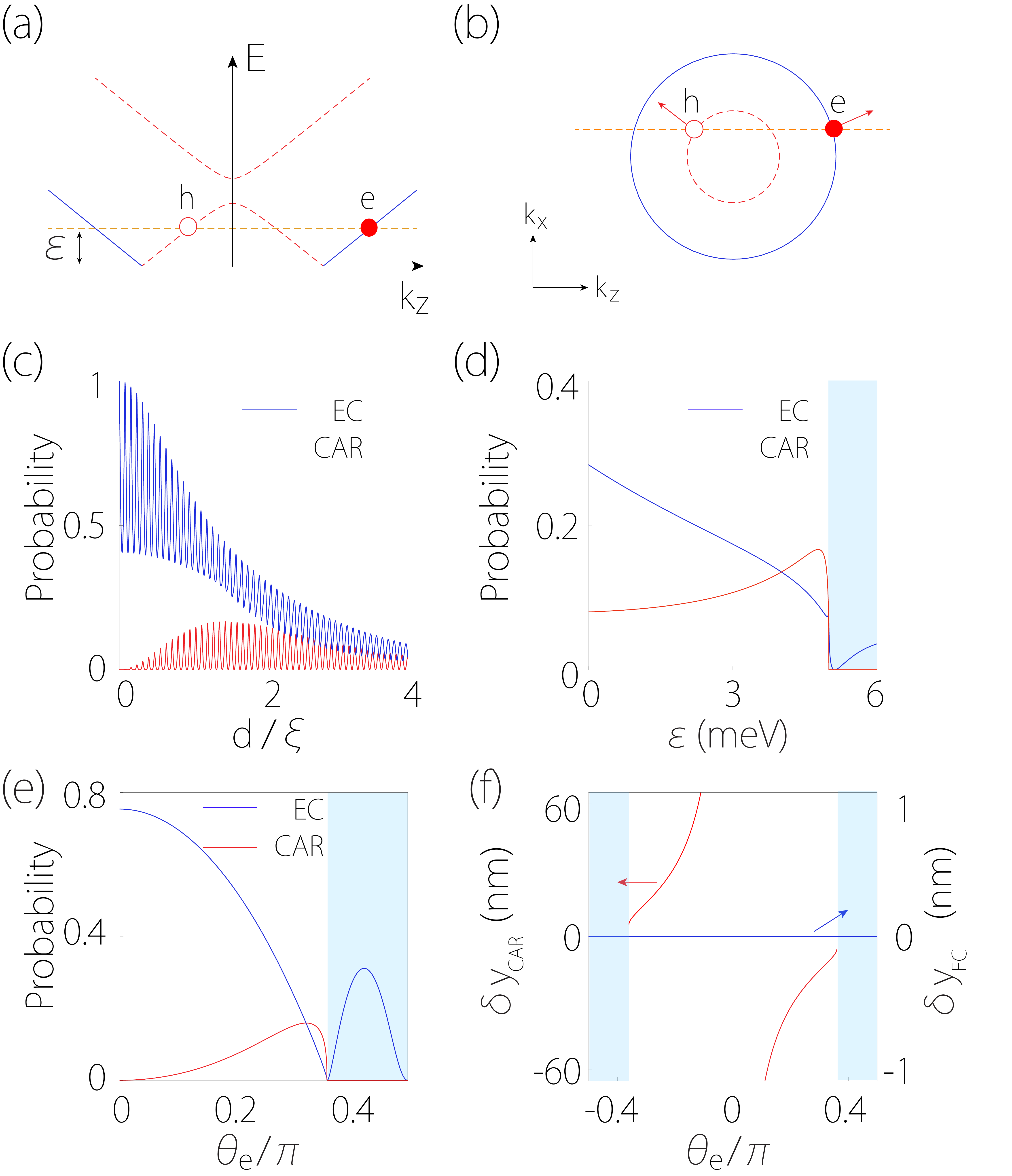}
	\caption{(a) BdG spectrum for the two N terminals in Model I at a fixed $k_x$ [marked by the dashed line in (b)]. The small gap appearing in the spectrum is due to the finite value of $k_x$. (b)
 Equi-energy contours at a fixed excitation energy $\varepsilon$. The solid (open) sphere denotes the incident electron (outgoing hole) state. The arrows in (b) denote the spin polarization directions.
 (c-e) Probabilities for CAR and EC versus (c) the width of the S region, (d) the excitation energy, and (e) the incident angle.  (f) Transverse shift for CAR (red) and EC (blue) versus the incident angle. The shaded regions in (d-f) mark the parameter ranges where CAR is forbidden.
 In the calculation, we choose $U_0=451$ meV, $E_F=40$ meV, $\Delta_0=5$ meV (coherence length $\xi=63$ nm), $v_x=v_y=v_z=1.5\times 10^6$ m/s. In (c) and (d), we set $\theta_e=7\pi/25$; in (d-f), we set $d=109.6$ nm; and in (c,e,f), we take $\varepsilon=2$ meV.  }
	\label{Fig2}
\end{figure}

In Fig.~\ref{Fig2}(a) and \ref{Fig2}(b), we show the schematic plot for the BdG spectrum and the equi-energy contours for the N region. The possible incoming electron and outgoing hole states are marked. Note that the transverse wave-vector component $\bm k_\|=(k_x,k_y)$, which is parallel to the interface, must be conserved during scattering. As a result, for a finite excitation energy $\varepsilon$, there exists a critical incident angle $\theta_c$, beyond which there is no propagating hole state for CAR (and also for local Andreev reflection). If the plane of incidence is the $xz$ plane, then the critical angle is given by
\begin{equation}\label{CA}
	\theta_c=\arctan\left(\frac{v_z|E_F-\varepsilon|}{2v_x\sqrt{E_F\varepsilon}}\right).
\end{equation}
Here, $E_F$ is measured from the Weyl point. For incident angle $|\theta_e|>\theta_c$, an incident electron from N1 cannot be transmitted as a hole in N2, and CAR does not occur.

We proceed to solve the scattering states for the BdG equation. Under $\mathcal{T}$, the electrons in the $\bm K_0$ valley is coupled to the holes in the $-\bm K_0$ valley, so the eigenstate has a four-component spinor form $\psi\equiv(\psi_{\bm K_0,\uparrow},\psi_{\bm K_0,\downarrow},\psi_{-\bm K_0,\downarrow}^*,-\psi_{-\bm K_0,\uparrow}^*)^T$, where the first two components are the electron spinor in the $\bm K_0$ valley, and the latter two are the hole spinor in the $-\bm K_0$ valley.

A scattering state takes the following form in each region:{{
\begin{eqnarray}
  \psi_{\bm k}(\bm r)=\left\{
  \begin{split}
    \psi^{e+}_{\bm k}+r_e\psi^{e-}_{\bm k}+r_h\psi^{h-}_{\bm k}, \ \ \ &z<0&\\
    a\psi'^{+}_S+b \psi'^{-}_S+c \psi''^{+}_S+d \psi''^{+}_S, \ \ \ &0<z<d&\\
    t_e\psi^{e+}_{\bm k}+t_h\psi^{h+}_{\bm k}, \ \ \ &z>d&
  \end{split}
  \right. ,\nonumber\\
\end{eqnarray}}}
where $r_{e(h)}$ is the amplitude for the normal (Andreev) reflection, $\psi'^{\pm}_S$ and $\psi''^{\pm}_S$ are the basis states for the S region, and $a,\ b,\ c, \ d$ are the corresponding amplitudes.

The basis states for the N region (including both N1 and N2) can be explicitly written down as{{
\begin{equation}
	\psi^{e+}_{\bm k}=\frac{1}{\sqrt{1-\eta_e^2}}\left[\begin{array}{c}
	e^{-i\alpha/2}\\
	\eta_e e^{i\alpha/2}\\
	0\\
	0
	\end{array}\right]e^{ik_xx+ik_yy+ ik_ez},
\end{equation}
\begin{equation}
\psi^{e-}_{\bm k}=\frac{1}{\sqrt{1-\eta_e^2}}\left[\begin{array}{c}
\eta_ee^{-i\alpha/2}\\
e^{i\alpha/2}\\
0\\
0
\end{array}\right]e^{ik_xx+ik_yy- ik_ez},
\end{equation}
\begin{equation}
	\psi^{h+}_{\bm k}=\frac{1}{\sqrt{1-\eta_h^2}}\left[\begin{array}{c}
	0\\
	0\\
	e^{-i\alpha/2}\\
	\eta_h e^{i\alpha/2}
	\end{array}\right]e^{ik_xx+ik_yy- ik_hz},
\end{equation}
\begin{equation}
\psi^{h-}_{\bm k}=\frac{1}{\sqrt{1-\eta_h^2}}\left[\begin{array}{c}
0\\
0\\
\eta_h e^{-i\alpha/2}\\
e^{i\alpha/2}
\end{array}\right]e^{ik_xx+ik_yy+ ik_hz}.
\end{equation}}}
Here, we have $\eta_{e(h)}=\chi\mathrm{sgn}(E_F\pm\varepsilon)\sqrt{\frac{E_F\pm\varepsilon-\chi v_zk_{e(h)}}{E_F\pm\varepsilon+\chi v_z k_{e(h)}}}$, $k_{e(h)}=v_z^{-1}\mathrm{sgn}(E_F\pm\varepsilon)\sqrt{(E_F\pm\varepsilon)^2-v_x^2k_x^2-v_y^2k_y^2}$, and $\alpha=\arctan\left({v_yk_y}/{v_xk_x}\right)$. The normalization factor ${1}/{\sqrt{1-\eta_{e(h)}^2}}$ is added to ensure that every propagating state carries the same particle current. Similarly, the basis states for the S region can be written down. Their explicit expressions are presented in the Appendix.

The boundary conditions at the two interfaces are
{{
\begin{equation}\label{BC}
\psi_{\bm k}|_{0-}=\psi_{\bm k}|_{0+},\qquad \psi_{\bm k}|_{d-}=\psi_{\bm k}|_{d+},
\end{equation}
}}
from which the four scattering amplitudes $r_{e(h)}$, $t_{e(h)}$ can be solved. The calculation is straightforward, and the explicit results are given in the Appendix. It should be noted that at a given energy, the scattering amplitudes are functions of the incident wave vector component $\bm k_\|$ that is parallel to the interface.

As our focus is on the nonlocal processes CAR and EC, in Fig.~\ref{Fig2}(c), we plot the probabilities for the two processes ($T_{e(h)}=|t_{e(h)}|^2$)as functions of the width $d$ of the S region. One observes that both curves exhibit typical Fabry-P\'{e}rot type oscillations with $d$, resulting from the interference between scattering processes at the two interfaces. Ignoring the oscillation, the averaged values for the two decrease towards zero at large $d$, as expected. One observes that their values are still sizable at two to three times the coherence length, which is similar to the case of Dirac fermions in graphene~\cite{Cayssol2008}.
For small $d$ {{of}} the order of the coherence length, EC tends to dominate over CAR.
In Fig.~\ref{Fig2}(d) and \ref{Fig2}(e), we plot the two probabilities versus the excitation energy and the incident angle, respectively. One observes that for a fixed $d$, $T_h$ tends to reach the maximum at $\varepsilon\sim\Delta_0$, where it may dominate over the EC process.
With respect to the incident angle $\theta_e$, CAR is totally suppressed at perpendicular incidence ($\theta_e=0$), because then the transmitted hole has a spin opposite to that of the incident electron.

To calculate the transverse shift in CAR, we expand $t_h(\bm k')$ in the expression for $\bm \Psi^{h+}$ in Eq.~(\ref{Psih}) to first order around the central wave-vector $\bm k$~\cite{Jiang2015,Liu2017c}. Note that $\bm \Psi^{h+}$ has two nonzero spinor components $\Psi^{h+}_3$ and $\Psi^{h+}_4$. Assuming that $\bm k$ is in the $x$-$z$ plane, regarding the transverse shift which is in the $y$ direction, we have
\begin{equation}
  \Psi^{h+}_{3,4}\propto e^{-W_y^2\left[y\mp\frac{1}{2}\frac{\partial\alpha}{\partial k_y'}+\frac{\partial}{\partial k_y'}\mathrm{arg}(t_{h})\right]_{\bm k_\|}^2/2},
\end{equation}
where the two signs $\mp $ correspond to $\Psi^{h+}_3$ and $\Psi^{h+}_4$ respectively, and $\bm k_\|=(k_x,0)$ by the geometry that we specify.
 Here, $\alpha$ and $t_h$ are functions of $\bm k'$ which is the wave vector labelling the partial waves as in Eq.~(\ref{Psih}). 
This is to be compared with the incident beam with $\Psi^{e+}_{1,2}\propto e^{-W_y^2[y\mp \frac{1}{2}\frac{\partial\alpha}{\partial k_y'}]_{\bm k_\|}^2/2}$. Then, weighted by the nonzero spinor components for each beam, we can obtain the relative shift between the two along the transverse $(y)$ direction, given by
\begin{equation}
  \delta y_\text{CAR}=-\left[\frac{1}{2}\Big(\frac{1-\eta_e^2}{1+\eta_e^2}+\frac{1-\eta_h^2}{1+\eta_h^2}\Big)
  \frac{\partial\alpha}{\partial k_y'}+\frac{\partial}{\partial k_y'}\mathrm{arg}(t_h)\right]_{\bm k_\|}.
\end{equation}
In a similar way, the transverse shift in EC can also be obtained, which is
\begin{equation}
 \delta y_\text{EC}=-\frac{\partial}{\partial k_y'}\mathrm{arg}(t_e)\Big|_{\bm k_\|}.
\end{equation}

Now we substitute the expressions for $\alpha$ and $t_{h(e)}$ into the above formulas. After simplification, we find a nice result given by
\begin{eqnarray}\label{Shift1}
	\delta y_\text{CAR}&=&-\frac{\chi}{2}\frac{v_zv_y}{v_x}\left(\frac{\cot\theta_e}{E_F+\varepsilon}
+\frac{\cot\theta_h}{E_F-\varepsilon}\right),\nonumber\\
\delta y_\text{EC}&=&0.
\end{eqnarray}
Here, the angles $\theta_{e(h)}=\arctan({k_x}/{k_{e(h)}})$. One observes that for this model, the transverse shift is zero for EC, but it is nonzero for CAR.  Importantly, the shift in CAR depends on the chirality of the Weyl fermions. Furthermore, the result is surprising in that $\delta y_\text{CAR}$ has no dependence on parameters of the S region, such as the pair potential and the band energy offset.

To understand this striking feature and to give an intuitive physical picture for the transverse shift, we show that the result can be reproduced via a symmetry argument. We note that when $v_x=v_y$, the system has an emergent rotational symmetry such that
\begin{equation}
[\mathcal{H}_\text{BdG}, \hat{J}_z]=0,
\end{equation}
where $\mathcal{H}_\text{BdG}$ is the BdG Hamiltonian in Eq.~(\ref{BdG}), and
\begin{equation}
\hat J_z=(\hat{\bm r}\times\hat {\bm k})_z+\frac{1}{2}\tau_0\otimes\sigma_z,
\end{equation}
where $\tau_0$ is the identity matrix in the Nambu space. $\hat J_z$ represents the total angular momentum along $z$. When evaluated for the quasiparticle beam, we should have $J_z=\langle \bm \Psi|\hat J_z|\bm \Psi\rangle$ to be conserved during scattering. For this model,
\begin{equation}
J_z=({\bm r}\times {\bm k})_z+\frac{1}{2}(\bm n)_z,
\end{equation}
where $\bm n$ is the spin polarization direction. For incident as well as transmitted electrons, we have $\bm n^e=(v_xk_x, v_yk_y, v_zk_{e})/(E_F+\varepsilon)$; whereas for the transmitted hole, we have $\bm n^h=(v_xk_x, v_yk_y, v_zk_{h})/(E_F-\varepsilon)$. Clearly, the spin angular momentum changes during CAR [see Fig.~\ref{Fig2}(b)], which must require a change in the orbital part for compensation. It follows that there must be a shift perpendicular to the incident plane given by
\begin{equation}\label{sym}
	\delta y_\mathrm{CAR}=\frac{\chi}{2k_x}(n_{z}^h-n_{z}^e)=-\frac{\chi}{2}v_z\Big(\frac{\cot\theta_e}
{E_F+\varepsilon}+\frac{\cot\theta_h}{E_F-\varepsilon}\Big),
\end{equation}
which exactly recovers the result derived using the scattering approach when $v_x=v_y$. On the other hand, for EC, there is no change in the spin direction, hence the transverse shift $\delta y_\mathrm{EC}$ should vanish.

The symmetry argument also helps to reveal the role of SOC behind the effect. It is because the orbital motion is coupled to the spin that once the spin state is changed in scattering (here in CAR), the orbital motion of the quasiparticle must also be changed. This is also the reason why we choose the model with the WSM terminals: the Weyl fermions in a sense have the strongest SOC. In addition, when the symmetry argument holds~\footnote{Note that for the current model, even if $v_x\neq v_y$, the symmetry argument can still apply after doing a scaling of the coordinates: $(x', y')=(x\sqrt{v_y/v_x},y\sqrt{v_x/v_y})$.}, we see that the transverse shift would only depend on the initial and final spin states in CAR, which are determined by the two N terminals and by the energy and the transverse momentum conservation laws. Hence, the detailed ($z$) variation of the pair potential and the band energy offset, as well as the details of the S region (including the layer thickness) do not affect the shift. This makes the transverse shift a robust physical effect.

In Fig.~\ref{Fig2}(f), we plot $\delta y_\mathrm{CAR}$ and $\delta y_\mathrm{EC}$ as functions of the incident angle. $\delta y_\mathrm{EC}$ vanishes identically, while $\delta y_\mathrm{CAR}$ is an odd function of the incident angle. $\delta y_\mathrm{CAR}$ becomes divergently large when approaching the perpendicular incidence, due to the small value of $k_x$ in Eq.~(\ref{sym}). Physically, the shift cannot ¡°diverge¡±. There are two factors which regulate this diverging behavior. First, the probability for CAR is completely suppressed at perpendicular incidence, so the seemingly diverging shift at perpendicular incidence cannot manifest in the measurement. Second, due to the uncertainty principle, a confined beam must have a finite spread in the wave vector (and hence the incident angle) distribution for the partial waves. When approaching perpendicular incidence, the diverging behavior indicates that the different partial waves would scatter in drastically different ways, such that the scattered beam would no longer be confined and the shift would become ill-defined. Thus, the diverging shift for the perpendicular incidence would not occur in reality.

When $\theta_e$ approaches the critical angle $\theta_c$, $\delta y_\mathrm{CAR}$ does not vanish, but approaches a finite value. This nonzero value can be directly seen from the symmetry argument (because the spins for the incident and the scattered states are in different directions) and obtained e.g., from Eq.~(\ref{sym}) by setting $|\theta_e|=\theta_c$. Beyond the critical angle, $\delta y_\mathrm{CAR}$ is no longer defined, because the CAR process does not occur in that regime.

Comparing Figs.~\ref{Fig2}(e) and \ref{Fig2}(f), one observes that when $\theta_e$ is small, the shift for CAR is large but the probability for CAR is small; while at large $\theta_e$,  the probability becomes large but the shift is small. This kind of behavior can be explained by noticing the two quantities' different dependence on the change in spin state during scattering. Nevertheless, both can be sizable in the intermediate range, where
$\delta y_\mathrm{CAR}$ can reach tens of nm. Here, the plot is made for carriers with positive chirality (left-handed). The result for negative chirality would have a reversed sign. The drastic difference between $\delta y_\mathrm{EC}$ and $\delta y_\mathrm{CAR}$ would make it possible to spatially separate the transmitted electrons and holes, which we will further discuss in Sec.~VI.

\section{Model II: SOC-Metal/S/SOC-Metal Junction}

In the last section, we have demonstrated the existence of a sizable transverse shift in CAR using a junction model based on WSMs. It leaves the following questions to be addressed. (i) Is the presence of Weyl points necessary for a finite shift? More generally, does the shift require any form of band crossing points? (ii) We have pointed out the importance of SOC in the N terminals, but how about the S region? Is it necessary to have SOC in S? Part of the answers can already be inferred from the symmetry argument we have presented, however, it is more desirable to have an explicit demonstration.

To this end, we consider the following model. The two N terminals are assumed to be identical and described by a model for a spin-orbit-coupled metal (SOC-metal), with
\begin{equation}\label{H0}
	H_0=\frac{1}{2 m_\text{N}}(-\nabla^2+M)\sigma_z-iv\sigma_x\partial_x-iv\sigma_y\partial_y,
\end{equation}
where $m_\text{N}$ is the effective mass. This model has the advantage that it can describe two distinct phases depending on the value of the parameter $M$. If $M<0$, it is a WSM with one pair of Weyl points on the $k_z$ axis at $k_z=\pm\sqrt{-M}$. These two Weyl points are of opposite chirality, hence the model breaks $\mathcal{T}$. On the other hand, if $M>0$, the conduction band and the valence band are fully separated by a gap, and it is metallic when $E_F>M/2m_\text{N}$ (for $E_F$ close to the conduction band bottom, it represents a doped semiconductor). The energy spectra for the two phases are schematically illustrated in Fig.~\ref{Fig3}(a) and Fig.~\ref{Fig4}(a), respectively.

Regarding the S region, we use the simplest quadratic model without SOC. The corresponding BdG Hamiltonian takes the form of
\begin{equation}\label{HS}	
\mathcal{H}_\text{S}=\left[\left(-\frac{1}{2m_\text{S}}\nabla^2-U_0-E_F\right)\tau_z+\Delta_0\tau_x\right]\otimes\sigma_0,
\end{equation}
where $m_\text{S}$ is the effective mass in S, $\tau_i$'s are the Pauli matrices acting on Nambu space, and $\sigma_0$ is the identity matrix in spin space.

In the following, we consider the two phases one by one. For each case, we investigate using the two approaches discussed in Sec.~III. For the scattering approach, fully analytical results are difficult to obtain for this model, so we proceed with numerical solutions.

\subsection{Case with $M<0$}

\begin{figure}[t]
	\includegraphics[width=8.4 cm]{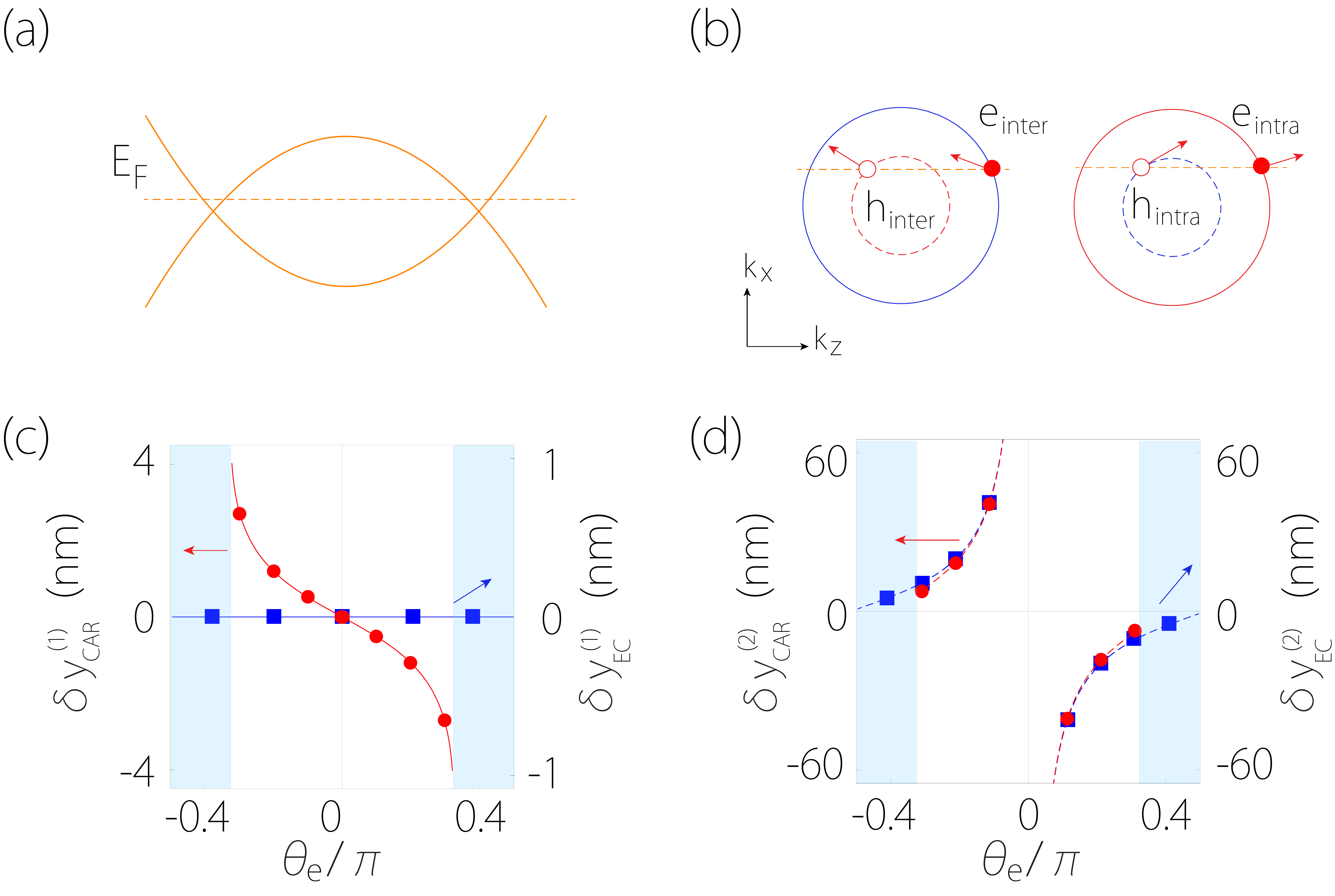}
	\caption{(a) BdG spectrum and (b) its equi-energy contours for Model II when $M<0$. The solid (open) sphere denotes the relevant electron (hole) states, assuming the incident electron corresponds to the rightmost state marked in (b). There exist both intravalley and intervalley scattering processes for transmission. The arrows represent the spin polarization directions for the states.
(c,d) Transverse shifts versus the incident angle for (c) the two intravalley processes and (d) the two intervalley processes.  In (c) and (d), the curves are obtained from the symmetry argument, while the data points are from the numerical solution using the scattering approach. Here, we take $U_0=154$ meV, $E_F=40$ meV, $v=1.5\times10^6$ m/s, $\Delta_0=5$ meV, $\varepsilon=3$ meV, $m_\text{N}=0.05m_e$ ($m_e$ is the free electron mass), $M/m_\text{N}=-0.5 $ eV, and $d=25$ nm.}
	\label{Fig3}
\end{figure}

As we have mentioned, when $M<0$, the model for {{N terminals has}} two Weyl points located at $k_z=\pm\sqrt{-M}$. At low energy, with $|E_F+\varepsilon|\ll |M/2m_N|$, the quasiparticles are described by the Weyl model
\begin{eqnarray}\label{WH}
  H_\pm=-iv\sigma_x\partial_x-iv\sigma_y\partial_y\mp iv_z\sigma_z\partial_z,
\end{eqnarray}
where $v_z=\sqrt{-M}/m_\text{N}$, and the subscript `$\pm$' refers to the two valleys (also corresponds to the chirality $\chi$). Hence, one may follow the similar procedure as in Sec.~III to do the calculation. However, two differences should be noted. First, here, the Weyl electron and its time reversal partner have opposite chiralities, due to the broken $\mathcal{T}$. This affects the change in spin state during scattering.
Second, when $\theta_e$ is small, there are four possible transmitted states: besides the intravalley scattering, there also exist two intervalley scattering processes [see Fig.~\ref{Fig3}(a,b)]. The scattering probabilities for the four processes depend on the detailed system parameters. Generally, if the two valleys are well separated in $k$ space and if the interfaces are not so sharp, the intravalley processes would be dominating, since the intervalley ones require a large momentum transfer.

In Fig.~\ref{Fig3}(c) and \ref{Fig3}(d), we plot the results for the transverse shifts in the four scattering processes. Here, the data points are calculated numerically using the full model in Eq.~(\ref{H0}). One observes that the shifts for the two intravalley processes are much smaller than those for the intervalley processes.

To understand this difference, we resort to the symmetry argument, for which we use the effective (Weyl) model in Eq.~(\ref{WH}). The spin directions for the scattered states are illustrated in Fig.~\ref{Fig3}(b). One can observe that the $z$-component of the spin has only small change for the intravalley processes (no change for intravalley EC), while it gets reversed for the intervalley processes. Thus, according to the symmetry argument, the transverse shift in intervalley processes should be larger. More explicitly, following similar analysis as in the previous section, we find that
\begin{eqnarray}
  \delta y^{(1)}_\mathrm{CAR}&=&\frac{\chi}{2}v_z\Big(\frac{\cot\theta_h}{E_F-\varepsilon}-\frac{\cot\theta_e}
  {E_F+\varepsilon}\Big),\\
  \delta y^{(1)}_\mathrm{EC}&=&0,
\end{eqnarray}
for the intravalley CAR and EC processes; while
\begin{eqnarray}
  \delta y^{(2)}_\mathrm{CAR}&=&-\frac{\chi}{2}v_z\Big(\frac{\cot\theta_h}{E_F-\varepsilon}+\frac{\cot\theta_e}
  {E_F+\varepsilon}\Big),\\
  \delta y^{(2)}_\mathrm{EC}&=&-\chi v_z\frac{\cot\theta_e}
  {E_F+\varepsilon},
\end{eqnarray}
for the intervalley processes. These results are plotted as solid and dashed curves in Fig.~\ref{Fig3}(c) and \ref{Fig3}(d). One observes that the results from symmetry argument agrees very well with the numerical results from the scattering approach based on the full model in Eq.~(\ref{H0}).

\begin{figure}[t]
	\includegraphics[width=8.4 cm]{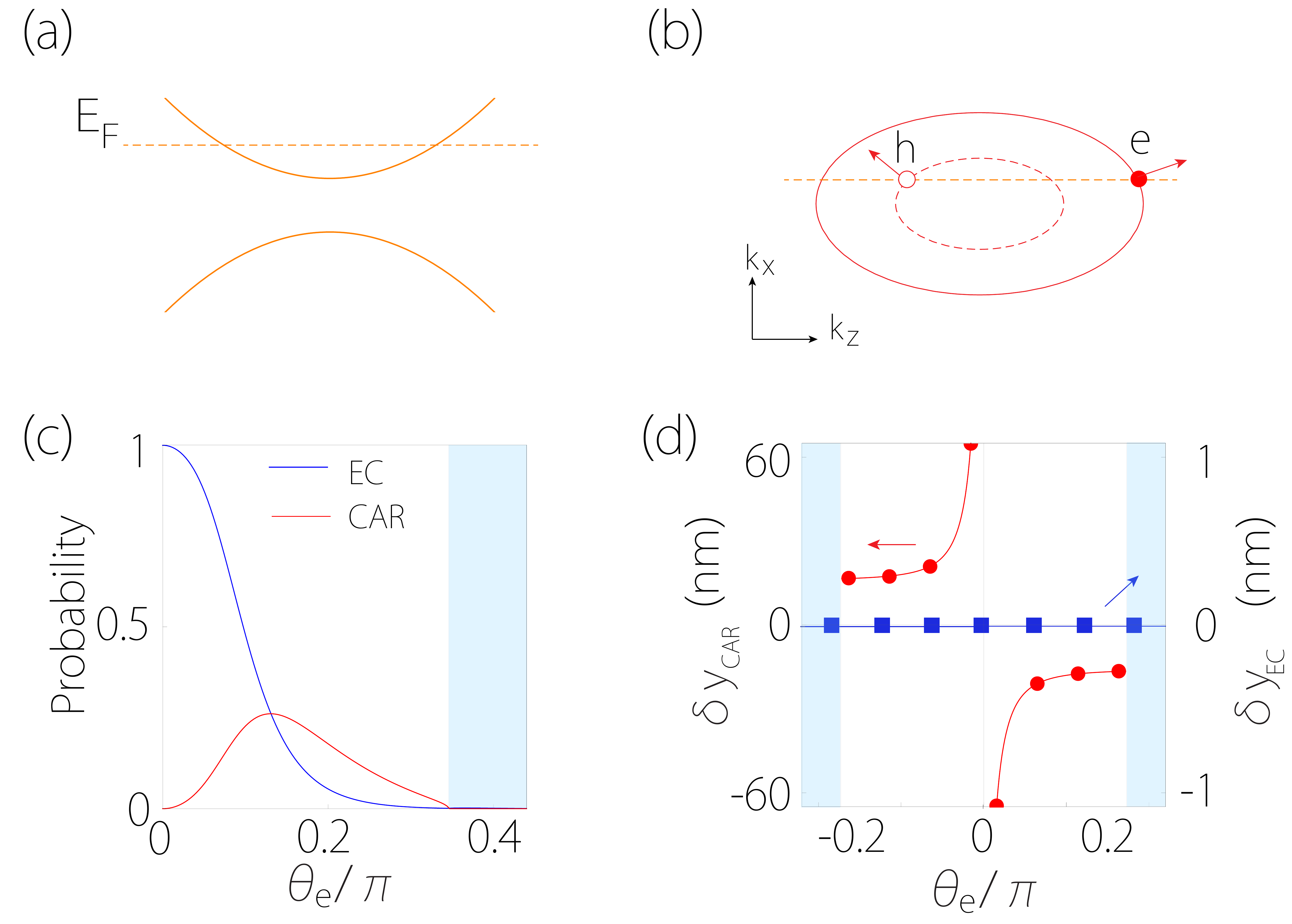}
	\caption{(a) BdG spectrum and (b) its equi-energy contours for Model II when $M>0$. The solid (open) sphere denotes the incident electron (outgoing hole) state. The arrows in (b) denote the spin polarization directions. (c) Probabilities for EC and CAR versus the incident angle. (d) Transverse shifts for CAR and EC. In (d), the solid curves are obtained from the symmetry argument, while the data points are from the numerical solution using the scattering approach.  Here, we set $\varepsilon=0.1$ meV, $U_0=118$ meV, $E_F=50$ meV, $m_\text{N}=0.05m_e$, $v=1.5\times10^6$ m/s, $\Delta_0=5$ meV, $M/m_\text{N}=80$ meV, and $d=40$ nm.}
	\label{Fig4}
\end{figure}

\subsection{Case with $M>0$}

When $M>0$, the two bands in the model Eq.~(\ref{H0}) are fully separated with a gap of $|M/m_\text{N}|$. Here, we consider the $n$-doped case, with $E_F>M/2m_\text{N}>0$. Then, there is a single Fermi surface for the N region, as illustrated in Fig.~\ref{Fig4}(b).

In Fig.~\ref{Fig4}(c), we plot the probabilities for CAR and EC as functions of $\theta_e$. One finds that EC dominates over CAR at small angles. At perpendicular incidence, CAR is completely suppressed, because the spin of the transmitted hole is opposite to that of the incident electron. Nevertheless, the probability for CAR can be sizable or even larger than EC at a finite $\theta_e$ [see Fig.~\ref{Fig4}(c)].

We have calculated the transverse shifts numerically via the scattering approach. The results are plotted as the data points in Fig.~\ref{Fig4}(d). One observes that while the shift in EC is zero, the shift in CAR has a finite value and becomes divergently large when $\theta_e\rightarrow 0$. Again, since the model preserves the rotational symmetry along $z$, we can derive the shift using the symmetry argument. For CAR, we have
\begin{equation}\label{CARs}
\delta y_\mathrm{CAR}=\frac{1}{2k_x}(n_{z}^h-n_{z}^e),
\end{equation}
with
\begin{equation}
n_{z}^{e/h}=\pm\frac{[(E_F\pm\varepsilon)^2-v^2k_x^2]^{1/2}}{E_F\pm \varepsilon}.
\end{equation}
For EC, one finds that $n_z$ does not change in the process, thus
\begin{equation}\label{ECs}
\delta y_\mathrm{EC}=0.
\end{equation}
In Fig.~\ref{Fig4}(d), we plot the results (\ref{CARs}) and (\ref{ECs}) by solid curves. One observes that they agree perfectly with the numerical results obtained from the scattering approach.

Based on the above results, we can now answer the questions raised at the beginning of this section. More specifically, we have demonstrated the following points.  First, the Weyl point or any type of band crossing (in either N or S) is not necessary for the existence of transverse shift in CAR. Second, in the first and the second models, SOC in N is crucial, but it is not necessary to have SOC for the S region (although it does affect the probabilities for the scattering processes). Here, S only plays the role as a channel for the electron-hole conversion. In addition, in the second model with $M>0$, the transmitted electrons do not have a transverse shift, while the transmitted holes have a finite shift. Around a finite incident angle, the shift has a definite sign, which would lead to a charge flow in the transverse direction.

\section{Model III: p/S/n Junction}

\begin{figure}[t]
	\includegraphics[width=8.4 cm]{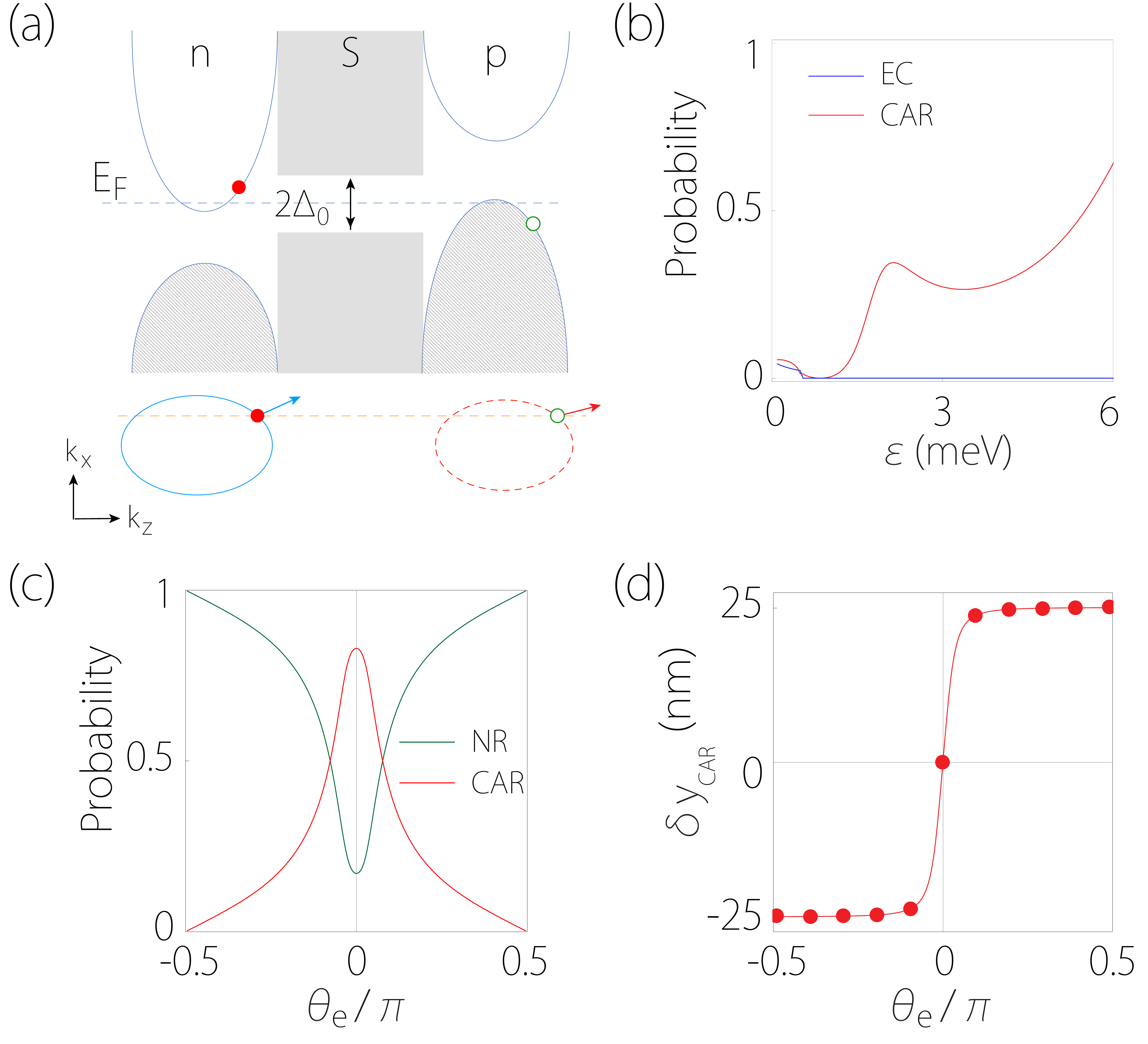}
	\caption{(a) Schematic energy diagram for a $p$S$n$ junction. The lower panel shows the equi-energy contours at an excitation energy where only electron states exist for N1 and only hole states exist for N2. (b) Probabilities for CAR and EC versus the excitation energy. (c) Probabilities for CAR and normal reflection (NR) versus the incident angle. (d) Transverse shift for CAR versus the incident angle.  Here, we take $d=30$ nm, $m_N=0.05 m_e$, $E_F=40$ meV, $v=1.5\times 10^6$ m/s, $M/m_\text{N}=7.2 $ meV, $\Delta_0=5$ meV, $U_0=261$ meV, $U_n=37$ meV, and $U_p=44$ meV. In (b), we take $\theta_e=\pi/8$. In (c) and (d), we take $\varepsilon=2$ meV.  }
	\label{Fig3c}
\end{figure}

In the previous two models and in most NSN structures, the EC process makes a non-negligible contribution to the nonlocal transport, although in some small parameter ranges CAR may become dominant. As we mentioned in the Introduction, there has been continuous effort in enhancing the CAR contribution and suppressing the EC. One simple proposal was put forward by Veldhorst and Brinkman~\cite{Veldhorst2010}, in which the EC is suppressed by the energy filtering enforced by the band structure of the two terminals. This is achieved by making
one N terminal a $p$-type semiconductor and the other N terminal an $n$-type semiconductor. The corresponding structure is termed as a ``$p$S$n$" junction. As illustrated in Fig.~\ref{Fig3c}(a), such band alignment allows only CAR contribution in N2, when the excitation energy is beyond the band edge. In the following, we investigate whether the transverse shift can still exist for a $p$S$n$ junction.

Here, we adopt the SOC-metal model in Eq.~(\ref{H0}). The $p$- and $n$-type doping for the two terminals can be described by $U(\bm r)$, which now takes the profile of
\begin{eqnarray}
  U(\bm r)=\left\{
  \begin{split}
    U_n, \ \ \ &z<0&\\
   { -U_0}, \ \ \ &0<z<d&\\
    U_p, \ \ \ &z>d&
  \end{split}
  \right., \nonumber\\
\end{eqnarray}
such that $(M/2m_\text{N}+U_n)<E_F<(-M/2m_\text{N}+U_p)$. Then, EC is completely suppressed when $\varepsilon>(-M/2m_\text{N}+U_p)-E_F$.

In Fig.~\ref{Fig3c}(b), we plot the probabilities for CAR and EC as functions of $\varepsilon$, which indeed shows that EC becomes completely suppressed and only CAR exists when $\varepsilon$ is above a threshold value. Figure~\ref{Fig3c}(c) shows the dependence of the CAR probability on the incident angle in the regime where EC vanishes. For this model, the probability is maximum at perpendicular incidence. It is also noted that when $\varepsilon>E_F-(M/2m_\text{N}+U_n)$, the local Andreev reflection would also be suppressed, so only normal reflection and CAR are the possible processes and $|r_e|^2+|t_h|^2=1$. This is the case for Fig.~\ref{Fig3c}(c).

To calculate the transverse shift in CAR, we again use the two approaches discussed before. Via the quantum scattering approach, we have numerically calculated the shift. The result is plotted as the data points in Fig.~\ref{Fig3c}(d). Indeed, a finite transverse shift can still exist and is an odd function of the incident angle. The result can also be derived using the symmetry argument, since we still have the rotational symmetry along $z$. One readily finds that
\begin{eqnarray}\label{31}
  \delta y_\mathrm{CAR}=\frac{1}{2k_x}(n_z^{h2}-n_z^{e1}),
\end{eqnarray}
where
{{$n_z^{e1}=[(E_F-U_n+\varepsilon)^2-v^2k_x^2]^{1/2}/(E_F-U_n+\varepsilon)$, and $n_z^{h2}=[(E_F-U_p-\varepsilon)^2-v^2k_x^2]^{1/2}/(E_F-U_p-\varepsilon)$}}. The analytical formula is plotted as the solid curve in Fig.~\ref{Fig3c}(d), which agrees perfectly with the data points from numerical calculations. The symmetry argument also explains why the shift vanishes at perpendicular incidence. It is because the spin states for the incident and the CAR states are parallel, both pointing to the $+z$ direction.

\section{Discussion and Conclusion}

\begin{figure}[t]
	\includegraphics[width=8.4 cm]{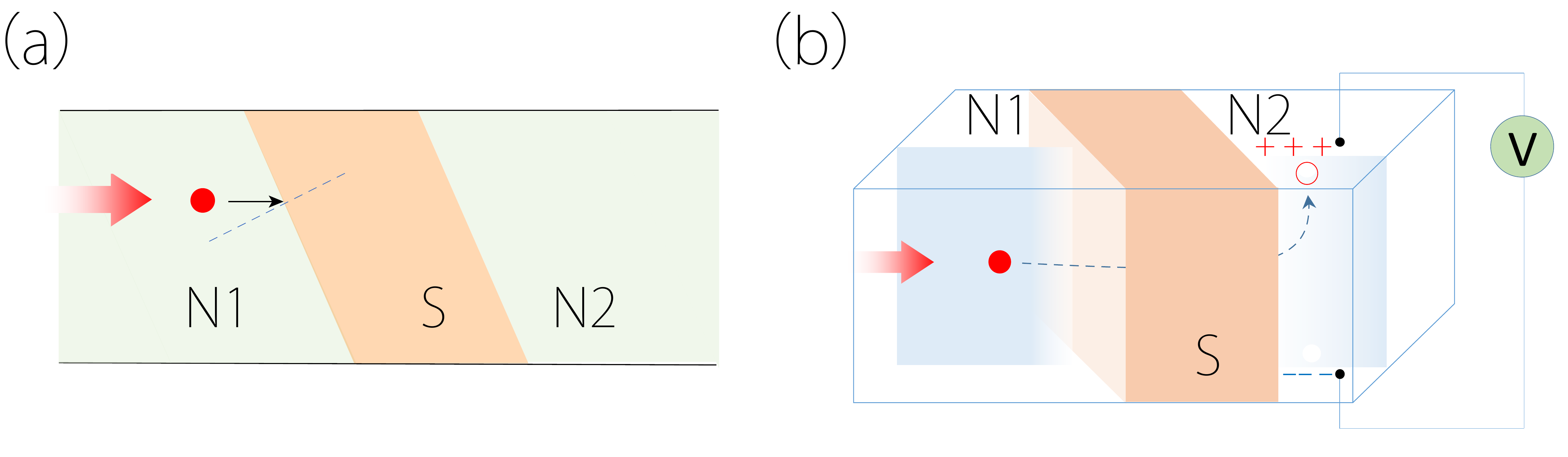}
	\caption{A possible setup for detecting the transverse shift in CAR.  (a) Top view of the junction. The electron flow is scattered at the interface with a finite average incident angle. (b) For systems described by Model II with $M>0$ or Model III (the $p$S$n$ junction), the transverse shift in CAR leads to
 a surface charge accumulation, which can be detected as a voltage difference between the top and bottom surfaces of N2 near the interface. }
	\label{Fig6}
\end{figure}

The main achievement of this work is that we reveal the general existence of a transverse spatial shift in the process of CAR. The shift can be sizable, with a magnitude much larger than the atomic scale, and it may exist for a wide range of parameters.

We have a few remarks before closing. First, we have obtained the same result via two different approaches. The first approach---the quantum scattering approach---is very general and applies without any constraint on the model parameters. The second approach based on symmetry argument, although applies only in the presence of the rotational symmetry,  offers a deep insight into the effect. Particularly, when the symmetry argument holds, the resulting shift only depends on the initial and final states in scattering,
independent of the details of the interfaces as well as the S region, leading to a universal type of behavior. For instance, the effects from possible interfacial barriers and/or variation of the pair potential beyond the step function model may affect the scattering probabilities, however, as long as they preserve the symmetry, the transverse shift for each scattering process will not be affected.

Second, the symmetry argument also makes it clear that the SOC in the N terminals are the key ingredient for the transverse shift studied here. The value of the shift depends on the SOC strength. For the Weyl model, the SOC strength is given by the Fermi velocities, which directly enters into the expression for the shift. For the SOC-metal model, the SOC strength is given by the coefficients for the three terms in Eq.~(\ref{H0}) (containing parameters $v$, $M$, and $m_N$).

Third, in this work, we have chosen the S layer to be a conventional $s$-wave superconductor. In principle, it may also be an unconventional superconductor with unconventional type of pair potential. In the case of local Andreev reflection, Yu \emph{et al.}~\cite{Yu2017} have shown that unconventional pairings can also induce anomalous transverse shifts with interesting features. Whether such shift also exists for CAR will be an interesting question to explore in future studies.

Fourth, for the first two models studied here, the shift in EC vanishes. This is due to the fact that the two N terminals are taken to be identical. If the two terminals are different, e.g., with different doping levels or with different materials, then there could in principle be a nonzero transverse shift also in EC. (For Weyl electrons, this is similar to the effect studied in Ref.~\cite{Jiang2015,Yang2015b}) Nevertheless, the value of the shift in EC should generally be different from that in CAR. This difference in the transverse shift would provide a possible way to spatially separate the transmitted holes and electrons.

Finally, for experimental detection, the most direct way is to engineer a collimated electron beam and inject it into the NSN structure just like in Fig.~\ref{Fig1}, then detect the shifted outgoing hole beam on the other side by using a collector. Although somewhat challenging, the technique for producing a collimated electron beam has actually been developed in the field of electron optics~\cite{Spector1990,Molenkamp1990,Dragoman1999}. Another more practical method is to use a setup as illustrated in Fig.~\ref{Fig6}. Here, the geometry for the NSN junction is designed such that the incident electrons that hit the NS interface have a finite average incident angle. Hence, the average shift for the outgoing holes in CAR have a definite sign. With such geometry, for the WSM/S/WSM model in Sec.~III, the transverse shift in CAR leads to {{a chirality accumulation on}} the top and bottom surfaces of N2, which can be detected by the imbalanced absorbance of the left and right circularly polarized light~\cite{Hosur2015,Liu2018}. For the SOC-metal model in Sec.~IV and the $p$S$n$ model in Sec.~V, the shift leads to a net charge accumulation on the surface, which can be electrically detected as a voltage signal as illustrated in Fig.~\ref{Fig6}(b). In these systems, the shift and the voltage signal are purely associated with CAR (note that the bulk anomalous Hall effect does not contribute
when the system has a twofold rotational axis along $z$), hence the effect also provides a possible all-electrical method for detecting CAR.

In conclusion, we discover the existence of a transverse spatial shift in CAR. We explicitly demonstrate the effect in three model systems. We show that the shift arises as a result of the SOC in the normal terminals. When there is an emergent rotational symmetry, the shift in CAR would have a universal behavior that it only depends on the initial and final states in CAR, independent of other system details. When the two N terminals are identical, the shift in EC vanishes, but the shift in CAR can be sizable. We propose possible setups for detecting the effect with optical or electrical signals. This also provides a promising alternative way for detecting CAR in experiment.

\begin{acknowledgements}
The authors thank Yee Sin Ang and D. L. Deng for valuable discussions. This work was supported by the Singapore MOE AcRF Tier 2 (MOE2017-T2-2-108), by NSFC (Grants No.~11574245, 11534001 and 11822407), by NBRPC (Grant No.~2014CB920901), and by NSF of Jiangsu Province, China (Grant No.~BK20160007).
\end{acknowledgements}

\begin{appendix}

\renewcommand{\theequation}{A\arabic{equation}}
\setcounter{equation}{0}
\renewcommand{\thefigure}{A\arabic{figure}}
\setcounter{figure}{0}
\renewcommand{\thetable}{A\arabic{table}}
\setcounter{table}{0}

\section{Analytic solution for scattering amplitudes for Model I}

The basis states for the S region in the WSM/S/WSM model can be obtained as
\begin{equation}
	\psi'^{+}_S=\left[\begin{array}{c}
	\chi e^{-i\alpha/2}\\
	\gamma e^{i\alpha/2}\\
	\chi e^{-i\alpha/2}e^{-i\chi \beta}\\
	\gamma e^{i\alpha/2}e^{i\chi\beta}
	\end{array}\right]e^{ik_xx+ik_yy+ik_0z-\kappa z},
\end{equation}
\begin{equation}
\psi'^{-}_S=\left[\begin{array}{c}
\gamma \chi e^{-i\alpha/2}\\
e^{i\alpha/2}\\
\gamma \chi e^{-i\alpha/2}e^{i\chi \beta}\\
e^{i\alpha/2}e^{-i\chi\beta}
\end{array}\right]e^{ik_xx+ik_yy-ik_0z+\kappa z},
\end{equation}
\begin{equation}
\psi''^{+}_S=\left[\begin{array}{c}
\chi e^{-i\alpha/2}\\
\gamma e^{i\alpha/2}\\
\chi e^{-i\alpha/2}e^{-i\chi \beta}\\
\gamma e^{i\alpha/2}e^{i\chi\beta}
\end{array}\right]e^{ik_xx+ik_yy+ik_0z+\kappa z},
\end{equation}
\begin{equation}
\psi''^{-}_S=\left[\begin{array}{c}
\gamma \chi e^{-i\alpha/2}\\
e^{i\alpha/2}\\
\gamma \chi e^{-i\alpha/2}e^{i\chi \beta}\\
e^{i\alpha/2}e^{-i\chi\beta}
\end{array}\right]e^{ik_xx+ik_yy-ik_0z-\kappa z}.
\end{equation}
Here we have $k_0=v_z^{-1}\sqrt{(E_F+U_0)^2-v_x^2k_x^2-v_y^2k_y^2}$, $\gamma=\sqrt{\frac{E_F+\varepsilon-\chi v_z k_0}{E_F+\varepsilon+\chi v_z k_0}}$, $\kappa=\Delta_0\sin\beta/v_z$. The parameter $\beta=\arccos(\varepsilon/\Delta_0)$, when $\varepsilon<\Delta_0$; and $\beta=-i\cosh^{-1}(\varepsilon/\Delta_0)$, when $\varepsilon>\Delta_0$. One observes that due to the superconducting pair potential, the basis states are mixtures with both electron and hole components.

Using the boundary conditions in Eq.~(\ref{BC}) to connect the wave function in the three regions, we can obtain the four scattering amplitudes. Straightforward calculation gives the following analytical results. {{
\begin{eqnarray}\label{AM}
  r_e&=&2\mathcal N^{-1}\Big[\eta_e(\Lambda^e_++i\Lambda^e_-)-\eta_h(1+\eta_e^2)\sinh^2(\kappa d)\Big],\nonumber\\
\end{eqnarray}
\begin{eqnarray}
   r_h&=&-2\mathcal N^{-1}\sinh(\kappa d)\sqrt{(1-\eta_e^2)(1-\eta_h^2)}(\Lambda_-^h+i\Lambda^h_+),\nonumber\\
\end{eqnarray}
where
$\Lambda^e_+\equiv (1+\eta_h^2)\{\cos^2\beta\sinh^2(\kappa d)+\sin^2\beta[\cos(2k_0 d)-\cosh^2(\kappa d)]\}$, $\Lambda^e_-\equiv (1-\eta_h^2)\sin\beta[\cos\beta\sinh(2\kappa d)+\sin\beta\sin(2k_0 d)]$, $\Lambda_+^h\equiv (1+\eta_e\eta_h)\sin\beta\cosh(\kappa d)$, $\Lambda_-^h\equiv (1-\eta_e\eta_h)\cos\beta\sinh(\kappa d)$.

The transmission amplitudes are given by
\begin{eqnarray}
t_e&=&\mathcal N^{-1}(1-\eta_e^2)\sin\beta e^{-ik_ed}(\Gamma_+^e+i\Gamma_-^e),
\end{eqnarray}
\begin{eqnarray}
   t_h&=&-2\mathcal N^{-1}e^{ik_hd}\sqrt{(1-\eta_e^2)(1-\eta_h^2)}\sin\beta\nonumber\\
  &&\times\sinh(\kappa d)(\Gamma_-^h+i\Gamma_+^h),
\end{eqnarray}}}
where $\Gamma_+^e\equiv (1+\eta_h^2)[e^{\kappa d}\cos(k_0 d+\beta)-e^{-\kappa d}\cos(k_0 d-\beta)]$, $\Gamma_-^e\equiv (1-\eta_h^2)[e^{\kappa d}\sin(k_0 d+\beta)-e^{-\kappa d}\sin(k_0 d-\beta)]$, $\Gamma_+^h\equiv (\eta_e+\eta_h)\cos(k_0d)$, and $\Gamma_-^h\equiv (\eta_e-\eta_h)\sin(k_0d)$. These functions are real when $\varepsilon<\Delta_0$.
The factor $\mathcal N$ is defined as $\mathcal N\equiv 4\eta_e\eta_h\sinh^2(\kappa d)-2(1+\eta_e^2\eta_h^2)[\cos^2\beta\sinh^2(\kappa d)-\sin^2\beta\cosh^2(\kappa d)]-2(\eta_e^2+\eta_h^2)\cos(2k_0 d)\sin^2\beta-i[(1-\eta_e^2\eta_h^2)\sin(2\beta)\sinh(2\kappa d)+2(\eta_e^2-\eta_h^2)\sin(2k_0 d)\sin^2\beta]$.

One can check that the above results satisfy the relation $|r_e|^2+|r_h|^2+|t_e|^2+|t_h|^2=1$, when $\varepsilon<\Delta_0$ and $|\theta_i|<\theta_c$, as required by the quasiparticle conservation.

\end{appendix}

\bibliography{NSN_ref}

%\bibliography{}

\end{document}